\documentclass[a4paper, 11pt]{article}
\usepackage[T1]{fontenc}
\usepackage{jheppub,graphicx,epsf,amssymb,amsbsy,amsfonts,amssymb,amsmath, empheq,mathrsfs,appendix,soul,physics}
\usepackage{hyperref}
\usepackage{amsmath, amssymb, mathtools}
\usepackage[utf8]{inputenc}
\usepackage{comment}
\usepackage{tikz}
\usepackage{cancel}
 
\def\be{\begin{equation}}
\def\ee{\end{equation}}

\def\bg{\bar{g}}
\def\beq{\begin{eqnarray}}\def\eeq{\end{eqnarray}}
\def\ba#1\ea{\begin{align}#1\end{align}}
\def\bg#1\eg{\begin{gather}#1\end{gather}}
\def\bm#1\em{\begin{multline}#1\end{multline}}
\def\bmd#1\emd{\begin{multlined}#1\end{multlined}}

\def\D{\Delta}

\def\p{\phi}

\def\t{\tau}

\def\({\left(}
\def\){\right)}
\def\[{\left[}
\def\]{\right]}

\def\t{\text}

\def\p{\partial}

\def\D{\Delta}

\pdfoutput=1

\title{Worldsheet CFT$_2$ and Celestial CFT$_2$ : An AdS$_3$-CFT$_2$ perspective}

\author{Shamik Banerjee$^{\,a,b}$, Nishant Gupta$^{\,a}$, Sagnik Misra$^{\,a,b}$, }
\affiliation[a]{National Institute of Science Education and Research (NISER), Bhubaneswar 752050, Odisha, India}
\affiliation[b]{Homi Bhabha National Institute, Training School Complex, Anushaktinagar, Mumbai 400094, India}
\emailAdd{banerjeeshamik.phy@gmail.com}
\emailAdd{nishantgupta.phy@gmail.com}
\emailAdd{sagnik.misra@niser.ac.in}

\abstract{Celestial CFT$_d$ is the putative dual of quantum gravity in asymptotically flat 
$(d+2)$ dimensional space time. We argue that a class of Celestial CFT$_d$ can be engineered via AdS$_{d+1}$-CFT$_d$ correspondence. Our argument is based on the observation that if we zoom in near the boundary of (Euclidean) AdS$_{d+1}$ then the conformal isometry group of EAdS$_{d+1}$, which is SO$(d+2,1)$, contracts to the Poincare group ISO$(d+1,1)$. This suggests that the near boundary scaling limit of a theory of \textit{conformal} gravity on EAdS$_{d+1}$ should be dual to a boundary CFT$_d$ with ISO$(d+1,1)$ symmetry. This dual CFT$_d$, since the symmetries match, is an example of a Celestial CFT$_d$. Similarly, if we have a \textit{non-conformal} theory of gravity on EAdS$_{d+1}$ then the near boundary scaling limit of such a theory is dual to a (boundary) Celestial CFT$_d$ with \textit{only} (SO$(d+1,1)$) Lorentz invariance. Celestial CFTs with only Lorentz invariance have been recently studied in the literature. Now following this logic we discuss, among other things, the near boundary scaling limit of the bosonic string theory on Euclidean AdS$_3$ in the presence of the NS-NS B field. The AdS$_3$ part of the worldsheet theory is free in this limit and has been studied in the literature in different contexts. This limit describes a ``long string'' which wraps the (Euclidean) AdS$_3$ boundary and it has been argued that the space-time CFT$_2$ which describes the radial fluctuations of a long string is a Liouville CFT. According to our proposal, the dual CFT$_2$ which describes the \textit{long string sector} is an example of a \textit{Celestial} CFT$_2$ with \textit{only} (SO$(3,1)$)Lorentz invariance. We do not get a full ISO$(3,1)$ invariant Celestial CFT$_2$ in this way because the string theory does not have target space conformal invariance. }

\begin{document}
\maketitle
\flushbottom

\section{Introduction}

\begin{figure}[h!]\label{fig1}
\begin{tikzpicture}
	\draw[thick] (-3.0,-3.0) rectangle (3.0,3.0);
	
	\node[align=right] at (-5.5,0) {\textbf{Set of all}\\\textbf{Celestial CFT$_d$}};
	
	\draw[->, thick] (-4.5,0) to[out=0, in=180] (-3.0,0.5);
	
	\fill[gray!30] (0,0) circle (2.2);
	\draw[thick] (0,0) circle (2.2);
	
	\node[align=center, font=\footnotesize] at (0,0) {All Celestial CFT$_d$ dual to\\Near-Boundary limit of \\ (string) theories on AdS$_{d+1}$};
\end{tikzpicture}
	\caption{The box represents the set of all Celestial CFT$_d$. The subset in grey includes only those Celestial CFT$_d$ which are dual to the near-boundary limit of (string) theories on AdS$_{d+1}$. }
	\label{fig1}	
\end{figure}
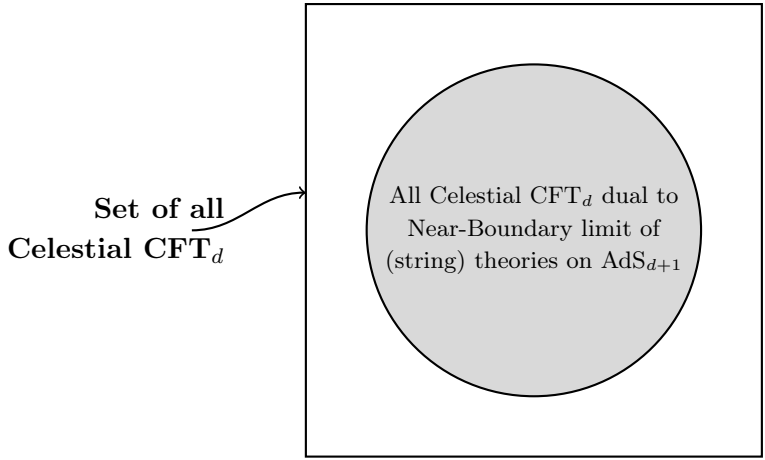

Celestial Holography \cite{Strominger:2017zoo,Pasterski:2021rjz,Donnay:2023mrd,Pasterski:2021raf,deBoer:2003vf,Pasterski:2016qvg,Pasterski:2017kqt,Banerjee:2019prz,Banerjee:2018gce} posits that the quantum theory of gravity in asymptotically flat $(d+2)$ dimensional space-time is dual to a (Celestial) conformal field theory living on the $d$ dimensional celestial sphere. The Lorentz group SO$(d+1,1)$ acts on the celestial sphere as the group of conformal transformations. The bulk space-time also has translation symmetry which is realized as internal symmetry in the Celestial CFT$_{d}$. So the global symmetry group of a Celestial CFT$_{d}$ is the Poincare group ISO$(d+1,1)$. However, developments \cite{Melton:2023bjw,Fan:2022vbz} in the last few years have taught us that sometimes it is more natural or convenient to work with Celestial CFTs with \textit{only} Lorentz invariance. In such Celestial CFTs correlation functions have the standard CFT form and there are specific algorithms following which one can reconstruct the Poincare invariant S-matrix elements from the Celestial CFT correlation functions. In interesting recent works the authors of \cite{Stieberger:2023fju,Melton:2024akx,Donnay:2025yoy} have constructed examples of such Celestial CFTs in two dimensions as Liouville theory coupled to conformal matter fields. 

In this paper we argue that a subset of Celestial CFTs can be studied via a scaling limit of AdS-CFT correspondence where we zoom in near the boundary of AdS. The main result is summarized in Fig-\eqref{fig1}. 

The paper is organized as follows: In sec-\eqref{2} we discuss the near boundary limit of the conformal isometry group of AdS$_{d+1}$. In sec-\eqref{3} we state our proposal relating the near boundary limit of a theory on AdS$_{d+1}$ to the boundary Celestial CFT$_d$. Sec-\eqref{3} onwards we discuss the near boundary limit of various theories on AdS and the emergent Carrollian structure. In particular, in sec-\eqref{7} we discuss the near boundary limit of the bosonic string theory on Euclidean AdS$_3$ in the presence of the NS-NS flux and discuss the relation between the (long strings or) worldsheet instantons which wrap the boundary S$^2$ and the dual \textit{Celestial} CFT$_2$ which lives on it. We also discuss the near boundary limit of Einstein gravity in the Appendix. 

\section{Zooming in near the boundary of Euclidean A\lowercase{d}S}\label{2}
	For the sake of concreteness let us consider EAdS$_3$ in Poincare coordinates. The line element is given by
	\begin{align}
	ds^2 =\frac{ d\eta^2+dz\,d\bar z}{\eta^2} \label{ds3_planar}
\end{align} 
The boundary is at $\eta=0$. The \textit{conformal} Killing vector fields (CKVs) of AdS$_3$ \eqref{ds3_planar} can be written in the following form\footnote{See for example \cite{Gupta:2022mdt}.}
	\begin{align}
		\mathcal L_{m}&=-z^{m+1}\,\partial_z +\,\frac{1}{2}\,m(m+1)\,\eta^2\,\partial_{\bar z}-\frac{1}{2}(m+1)\,z^m\,\eta\,\partial_{\eta} \cr
			\mathcal{\bar L}_{m}&=-\bar z^{m+1}\,\partial_{\bar z} + \,\frac{1}{2}\,m(m+1)\,\eta^2\,\partial_{ z}- \frac{1}{2}(m+1)\,\bar z^m\,\eta\,\partial_{\eta} \cr
			\mathcal P_{r,s}&=-z^{r+\frac{1}{2}}\,\bar z^{s+\frac{1}{2}}\,\partial_{\eta}+ 2\eta\,\left(s+\frac{1}{2}\right)\,z^{r+s}\partial_z+2\eta\,\left(r+\frac{1}{2}\right)\,\bar z^{r+s}\partial_{\bar z}\cr &+(r+\frac{1}{2})(s+\frac{1}{2})\,\eta^2\,\partial_{\eta} \label{so23_vec}
	\end{align}
where $m \in \{0,\pm 1\}$ and $(r,s)\in \{\pm \frac{1}{2}\}$. 
    
    These vector field obey commutation relations,
	\begin{align}\label{algebra1}
		&\left[\mathcal L_m,\mathcal L_n\right]=(m-n)\mathcal L_{m+n}\,,~~\left[\mathcal{\bar L}_m, \mathcal{\bar L}_n\right]=(m-n)\mathcal{\bar L}_{m+n}\cr
		&\left[\mathcal L_m, \mathcal P_{r,s}\right]=\frac{1}{2}\left(m-2r\right)\,\mathcal P_{m+r,s}\,,~\left[\mathcal{\bar L}_m, \mathcal P_{r,s}\right]=\frac{1}{2}\left(m-2s\right)\,\mathcal P_{r,m+s} \cr
		&\left[\mathcal P_{r,s}, \mathcal P_{r',s'}\right]=2\left(\epsilon_{rr'}\mathcal{\bar L}_{s+s'}+\epsilon_{ss'}\,\mathcal L_{r+r'}\right)
	\end{align}
where $\epsilon_{-\frac{1}{2}\frac{1}{2}}=-\epsilon_{\frac{1}{2}-\frac{1}{2}}=1$. The commutators \eqref{algebra1} is the Lie algebra of $SO(4,1)$ which is the group of conformal transformations of $\t{EAdS}_3$.

To zoom in near the boundary we define the new coorrdinate $\tilde{\eta}$ by \begin{align} 
	\eta= \epsilon \tilde \eta \label{coord_scaling}, \   \epsilon\rightarrow 0
	\end{align}
and take $\epsilon\rightarrow 0$ at fixed $\tilde\eta$. 

In terms of $\tilde{\eta}$ the CKVs become 
\begin{align}
		\mathcal L_{m}&=-z^{m+1}\,\partial_z-\frac{\tilde \eta}{2}(m+1)\,z^m\,\partial_{\tilde \eta} +\frac{\epsilon^2\tilde \eta^2}{2}\,m(m+1)\,\partial_{\bar z}\cr
	\mathcal{\bar L}_{m}&=-\bar z^{m+1}\,\partial_{\bar z}-\frac{\tilde \eta}{2}(m+1)\,\bar z^m\,\partial_{\tilde \eta}+\frac{\epsilon^2 \tilde \eta^2}{2}\,m(m+1)\,\partial_{ z} \cr
	\mathcal P_{r,s}&=-z^{r+\frac{1}{2}}\,\bar z^{s+\frac{1}{2}}\,\frac{\partial_{\tilde \eta}}{\epsilon}+\epsilon\,\bigg[2\,\tilde \eta\,\left(s+\frac{1}{2}\right)\,z^{r+s}\partial_z+2\,\tilde \eta\,\left(r+\frac{1}{2}\right)\,\bar z^{r+s}\partial_{\bar z}\cr &+(r+\frac{1}{2})(s+\frac{1}{2})\,\tilde \eta^2\,\partial_{\tilde \eta}\bigg].\label{so23epsilon}
\end{align}
Now we take $\epsilon \rightarrow 0$ keeping $\tilde \eta$ fixed. In this limit the vector fields $\mathcal P_{r,s}$ diverge and to cure this we introduce the rescaled vector fields $P_{r,s}$ defined as 
\be\label{trans}
 P_{r,s}= \epsilon\,\mathcal P_{r,s}, \  \epsilon\rightarrow 0
 \ee
So in the limit $\epsilon \rightarrow 0$ of \eqref{so23epsilon} we obtain the following vector fields,
	\begin{align}
				& L_{m}=-z^{m+1}\,\partial_z-\frac{1}{2}(m+1)\,z^m\,\tilde \eta\,\partial_{\tilde \eta}, \cr
			&\bar { L}_{m}=-\bar z^{m+1}\,\partial_{\bar z}-\frac{1}{2}(m+1)\,\bar z^m\,\tilde \eta\,\partial_{\tilde \eta} ,\cr
			& P_{r,s}=-z^{r+\frac{1}{2}}\,\bar z^{s+\frac{1}{2}}\,\partial_{\tilde \eta} ,\label{so23_flat}
	\end{align}
	where 
	\begin{align}\label{L}
		\mathcal L_{m} {\xrightarrow{\epsilon \rightarrow 0}}~~  L_{m}\,,~~\mathcal{\bar L}_{m} {\xrightarrow{\epsilon \rightarrow 0}}~~ \bar L_{m}.
	\end{align} 
	The vector fields \eqref{so23_flat} satisfy the $\t{ISO}(3,1)$ algebra,
	\begin{align}
		&\left[L_m, L_n\right]=(m-n) L_{m+n}\,,~~~~~~~~~\left[\bar{ L}_m, \bar {L}_n\right]=(m-n)\bar { L}_{m+n}\cr
		&\left[L_m, P_{r,s}\right]=\frac{1}{2}\left(m-2r\right)\, P_{m+r,s}\,,~~\left[\bar { L}_m, P_{r,s}\right]=\frac{1}{2}\left(m-2s\right)\, P_{r,m+s} \cr
		&~~~~~~~~~~~~~~~~~~~~~~~~~~~~~\left[ P_{r,s},  P_{r',s'}\right]=0 .\label{iso13}
	\end{align}

So we can see that in the near-boundary limit \eqref{coord_scaling} the conformal group of $\t{EAdS}_3$ contracts to the Poincare group $\t{ISO}(3,1)$. The parameter $\epsilon$ acts as the Inonu-Wigner contraction parameter. 

The Lorentz group SO$(3,1)\subset \t{ISO}(3,1)$ acts on the coordinates $(\tilde\eta, z, \bar z)$ as 
\be
\begin{gathered}
z \rightarrow \frac{az+b}{cz+d}, \  \bar z \rightarrow \frac{\bar a\bar z+\bar b}{\bar c\bar z+\bar d}, \  \tilde\eta \rightarrow \frac{\tilde\eta}{|cz+d|^2}
\end{gathered}
\ee
where we have used the isomorphism SO$(3,1)\rightarrow\t{SL}(2,\mathbb C)/\mathbb Z_2$ and 
\be
\begin{pmatrix}
a& b \\
c & d
\end{pmatrix}
\in \text{SL}(2,\mathbb C)
\ee
Similarly the translation $P_{r,s} \in\t{ISO}(3,1)$ acts as
\be
z \rightarrow z, \bar z \rightarrow \bar z, \tilde \eta \rightarrow \tilde\eta \ + \  a_{r,s} z^{r+\frac{1}{2}} \bar z^{s+\frac{1}{2}}
\ee
where $a_{r,s}=\bar a_{s,r}$.

Note that the Lorentz group SO$(3,1)$ is the isometry group of the degenerate metric 
\be\label{deg}
ds^2 = \frac{dz d\bar z}{\tilde\eta^2}
\ee
whereas the translations $P_{r,s}$ act as (Weyl) conformal transformations. So if we start with the AdS$_3$ isometry group then in the near boundary limit we will get only the Lorentz group SO$(3,1)$ acting as the isometries of the degenerate metric \eqref{deg}.

Our calculation can be generalized in variety of ways. For example, one could replace Euclidean $\t{AdS}_3$ by Lorentzian $\t{AdS}_3$ and take the same near-boundary limit \eqref{coord_scaling}. In this limit the conformal group $\t{SO}(3,2)$ of Lorentzian $\t{AdS}_3$ contracts to the "Poincare group" $\t{ISO}(2,2)$ of the split signature Minkowski space-time $\t{R}^{2,2}$. 

Similarly we can take the near boundary limit \eqref{coord_scaling} in any dimension and then the conformal group SO$(d+2,1)$ of EAdS$_{d+1}$ contracts to the Poincare group ISO$(d+1,1)$.

\section{Our Proposal}\label{3}

\begin{figure}[h!]
\centering
\begin{tikzpicture}
	\tikzstyle{box}=[draw, rectangle, minimum width=3cm, minimum height=1cm, text centered, align=center, thick]
	
	\node[box](box1) at (0,0) {Celestial CFT$_2$};
	\node[box](box2) at (9,0) {3-D (Carrollian) EFTs with gravity \\/ Long strings  };
	\node[box](box3) at (0,3) {Conventional CFT$_2$};
	\node[box](box4) at (9,3) {Gravitational EFT/String theory \\ on AdS$_3$ with NS-NS flux};
	
	\draw[<->, very thick] (box1) --
	node[above] {Duality} (box2);
	\draw[<->, very thick] (box3) -- node[above] {Duality} (box4);
	\draw[->, very thick] (box3) -- node[left] {??} (box1);
	\draw[->, very thick] (box4) -- node[left] {Near boundary  limit} (box2);
\end{tikzpicture}
\caption{Diagrammatic representation of our proposal}
\label{diagram}
\end{figure}
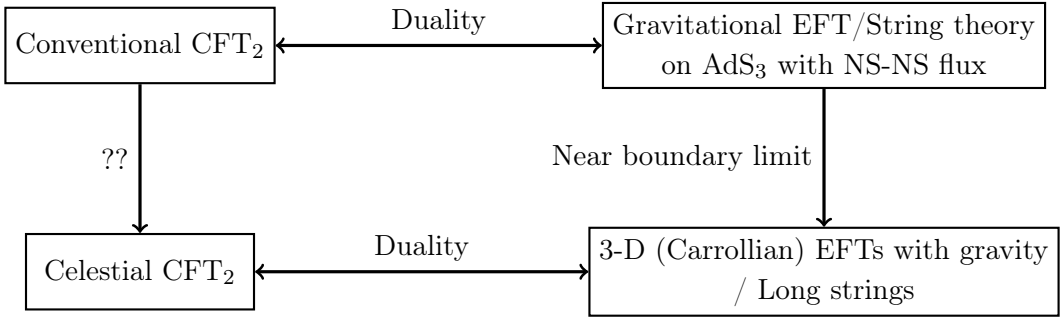

\subsection{Conformal gravity on AdS$_{d+1}$}

We have shown that in the near boundary limit \eqref{coord_scaling} the conformal group SO$(d+2,1)$ of EAdS$_{d+1}$\footnote{The E stands for Euclidean.} contracts to the Poincare group ISO$(d+1,1)$. Now AdS-CFT duality tells us that any theory of quantum gravity on asymptotically EAdS$_{d+1}$ space is holographically dual to a boundary (Euclidean) CFT$_{d}$. So if we have a theory of \textit{conformal} gravity coupled to conformal matter on EAdS$_{d+1}$ then that theory is also dual to a  CFT$_{d}$ with a \textit{larger}\footnote{The conformal symmetry group of CFT$_{d}$ is SO$(d+1,1)\subset \t{SO}(d+2,1)$.} symmetry group SO$(d+2,1)$ which is the same as the conformal isometry group of EAdS$_{d+1}$. Now, assuming that the duality survives the near boundary scaling limit, we can say that in this limit the SO$(d+2,1)$ symmetry of the dual CFT$_{d}$ also reduces to the Poincare group ISO$(d+1,1)$. Note that in the process the boundary conformal group SO$(d+1,1)\subset \t{ISO}(d,1)$ remains untouched. \\

So in a nutshell, \textit{the near boundary limit \eqref{coord_scaling} of any conformal gravity theory on EAdS$_{d+1}$ is holographically dual to a boundary} CFT$_{d}$ \textit{with symmetry} ISO$(d+1,1)$. So the dual is a \textit{Celestial} CFT$_d$. The same reasoning is true for conformal gravity theories on Lorentzian AdS$_{d+1}$ except that now in the near boundary limit the Celestial CFT$_d$ has symmetry group ISO$(d,2)$.  \\

This identification is based on symmetries alone. In other words, the correlation functions of the Celestial CFT$_d$ which arises in this way need not necessarily compute the  \textit{semiclassical} scattering amplitudes in \textit{asymptotically flat} $(d+2)$ dimensional space-time. 

Our proposal gives a way to study a \textit{subset} of celestial CFTs via AdS-CFT duality. In particular, this suggests that Celestial CFT$_2$ with symmetry group ISO$(3,1)$ arises in a certain scaling limit of the AdS$_3$-CFT$_2$ correspondence.  

\subsection{Einstein gravity on AdS$_{d+1}$}

Now suppose that the bulk EAdS$_{d+1}$ theory is matter coupled to Einstein gravity. Then the dual theory is a CFT$_{d}$ with symmetry SO$(d+1,1)$. We can take the near boundary scaling limit \eqref{coord_scaling} even in this case. Scaling limit does not change the bulk isometry algebra SO$(d+1,1)$ but it acts non-trivially on the correlation functions of bulk fields. Therefore in this limit the bulk theory changes. So again if we assume that the duality survives the scaling limit then we can say that the boundary CFT$_{d}$ changes to a \textit{Celestial} CFT$_{d}$ but now with \textit{only} (Lorentz) SO$(d+1,1)$ invariance. We do not get the full Poincare invariance because the EAdS$_{d+1}$ bulk theory we start with does not have the conformal group SO$(d+2,1)$ as (asymptotic) symmetry. \\

Therefore we can say that \textit{the near boundary limit \eqref{coord_scaling} of Einstein gravity coupled to matter on EAdS$_{d+1}$ is holographically dual to a Celestial CFT$_{d}$ with only (Lorentz) SO$(d+1,1)$ invariance.} Similarly, if we start with theories on Lorentzian AdS$_{d+1}$ then the dual Celestial CFT$_{d+1}$ has symmetry group SO$(d,2)$ which is the Lorentz group of the $(d+2)$ dimensional split signature Minkowski space. \\

Our proposal is that in the near boundary limit bulk theories of gravity on AdS should be dual to Celestial CFTs. In the following sections we discuss the near boundary limit of various theories on AdS including the worldsheet theory of bosonic strings on AdS$_3$ in the presence of NS-NS flux. In the appendix, we also discuss the near boundary limit of Einstein gravity on EAdS$_{d+1}$. 

\subsection{FAQ}
Is any (conventional) CFT$_d$ also a Celestial CFT$_d$ without bulk translation invariance, i.e, without the full ISO$(d+1,1)$ symmetry? \\

The answer is no but, we do not yet have the criteria to decide whether a given CFT is Celestial or not which depends only on the intrinsic properties of a CFT. Our proposal sheds light on why the answer is no in the cases of Holographic CFTs. We say a few words on this very important point now. \\

Given a CFT$_d$, which has a dual (semiclassical) gravity description on AdS$_{d+1}$, we can proceed in the following way. We take the near boundary limit of the dual AdS$_{d+1}$ theory and, according to our proposal, the resulting theory is dual to a Celestial CFT$_d$ living on the boundary. So the Celestial CFT$_d$ we end up with cannot be the same as the conventional (holographic) CFT$_d$ we started with. Currently we do not understand this limit in the field theory side. If we understand this then we will be able to take the ``Celestial limit'' of any (conventional) CFT without using the AdS-CFT duality. We leave this question for future research. 

\section{Near-Boundary limit of a CFT two-point function and emergent Carrollian structure}

Let us consider an Euclidean CFT$_{d+1}$ on AdS$_{d+1}$ with metric
\be\label{Poincare}
ds^2 = \frac{d\eta^2 + d\vec x^2}{\eta^2}, \   \vec x \in R^d
\ee 
Since \eqref{Poincare} is conformally flat the two-point function of a scalar conformal primary $\Phi_{\D}$ of dimension $\D$  is given by,
\begin{align}
	&\langle\Phi_{\D}({\eta_1}, \vec x_1)\,\Phi_{\D}({\eta}_2, \vec x_2)\rangle_{AdS_{d+1}}= \eta_1^{\Delta}\,\eta_2^{\Delta}\langle\Phi_{\D}({\eta_1}, \vec x_1)\,\Phi_{\D}({\eta}_2, \vec x_2)\rangle_{R^{d+1}} \label{2 pointdS3}
\end{align}
where $\langle\Phi_{\D}({\eta_1}, \vec x_1)\,\Phi_{\D}({\eta}_2, \vec x_2)\rangle_{\t{R}^{d+1}} $ is the two-point function of the scalar primary $\Phi_{\D}$ on $\t{R}^{d+1}$. Therefore,
\begin{align}
	&\langle\Phi_{\D}({\eta_1},z_1,\bar z_1)\,\Phi_{\D}({\eta}_2,z_2, \bar z_2)\rangle_{AdS_{d+1}}=\frac{\eta_1^{\Delta}\,\eta_2^{\Delta}}{\left(\left(\eta_{12}\right)^2+ \vec x_{12}^2\right)^{\Delta}}  \label{dS3_corr}
\end{align}
where $\vec x_{ij}=\vec x_i- \vec x_j$. 

To take the near-boundary limit of \eqref{dS3_corr} we change coordinate $\eta\rightarrow \tilde\eta=\frac{\eta}{\epsilon}$ and the expression for the two-point function becomes 
\begin{align}
	&\langle\Phi_{\D}(\epsilon\,{\tilde \eta_1}, \vec x_1)\,\Phi_{\D}(\epsilon\, \tilde \eta_2, \vec x_2)\rangle_{AdS_{d+1}}=\frac{\epsilon^{2\D}\tilde \eta_1^{\Delta}\,\tilde \eta_2^{\Delta}}{\left(\epsilon^2 \tilde\eta_{12}^2+ \vec x_{12}^2\right)^{\Delta}}. \label{2ptscalar}
\end{align}
Now we let $\epsilon\rightarrow 0$ keeping $\tilde\eta$ fixed. In this limit the correlation function can be expanded as\footnote{This type of formula appears for example in \cite{Pasterski:2017kqt}.}, 
\be\label{near}
\begin{gathered}
\langle\Phi_{\D}(\epsilon\,{\tilde \eta_1}, \vec x_1)\,\Phi_{\D}(\epsilon\, \tilde \eta_2, \vec x_2)\rangle_{AdS_{d+1}} \\
\rightarrow \epsilon^d \( \pi^{\frac{d}{2}} \frac{\Gamma\(\D- \frac{d}{2}\)}{\Gamma (\D)} \frac{\tilde\eta_1^{\D}\tilde\eta_2^{\D}}{|\tilde\eta_{12}|^{2\D-d}} \delta^{d}(\vec x_1 - \vec x_2)\) + \epsilon^{2\D} \(\frac{\tilde\eta_1^{\D}\tilde\eta_2^{\D}}{|\vec x_1 - \vec x_2|^{2\D}}\)
\end{gathered}
\ee
So we consider two different cases depending on the dimension $\D$: 

\subsection{Electric Sector : \textbf{$\D >  \frac{d}{2}$}}  In this case the first term in \eqref{near} is dominant as $\epsilon\rightarrow 0$. So if $\D > \frac{d}{2}$ then the near boundary limit of the two point function is 
\be
\langle\Phi_{\D}(\epsilon\,{\tilde \eta_1}, \vec x_1)\,\Phi_{\D}(\epsilon\, \tilde \eta_2, \vec x_2)\rangle_{AdS_{d+1}} \\
\rightarrow \epsilon^d \( \pi^{\frac{d}{2}} \frac{\Gamma\(\D- \frac{d}{2}\)}{\Gamma (\D)} \frac{\tilde\eta_1^{\D}\tilde\eta_2^{\D}}{|\tilde\eta_{12}|^{2\D-d}} \delta^{d}(\vec x_1 - \vec x_2)\), \ \D> \frac{d}{2}
\ee
We can define rescaled fields as 
\be\label{elec}
\Phi_{\D}(\epsilon\,{\tilde \eta}, \vec x) = \epsilon^{\frac{d}{2}} \tilde\Phi_{\D}({\tilde \eta}, \vec x)
\ee
in terms of which the correlation function becomes 
\be\label{electric}
\langle\tilde\Phi_{\D}({\tilde \eta_1}, \vec x_1)\,\tilde\Phi_{\D}(\tilde \eta_2, \vec x_2)\rangle_{\t{Near-Boundary}} = 
  \pi^{\frac{d}{2}} \frac{\Gamma\(\D- \frac{d}{2}\)}{\Gamma (\D)} \frac{\tilde\eta_1^{\D}\tilde\eta_2^{\D}}{|\tilde\eta_{12}|^{2\D-d}} \delta^{d}(\vec x_1 - \vec x_2), \ \D> \frac{d}{2}
\ee
This has precisely the structure of the two point function of Carrollian \cite{levy,SenGupta:1966qer} conformal primaries \cite{Banerjee:2018gce}  in the electric sector a Carrollian CFT$_{d+1}$ living on the degenerate metric 
\be
ds^2 = \frac{d\vec x^2}{\tilde\eta^2}, \   \vec x \in R^d
\ee 
Note that the conformal group of this metric which is ISO$(d+1,1)$ is isomorphic to the conformal Carroll group \cite{Duval:2014lpa} in $(d+1)$ dimensions. So this result is not unexpected. We can get the more familiar form of the Electric sector two point function if we do a conformal rescaling and define 
\be\label{flat}
\Phi_{\D}(\tilde\eta, \vec x) = \epsilon^{\frac{d}{2}} \tilde\Phi_{\D}(\tilde\eta, \vec x) = \epsilon^{\frac{d}{2}}\tilde\eta^{\D}\tilde\Phi_{\D}'(\tilde\eta, \vec x) 
\ee
In terms of the primed fields the two point function becomes 
\be\label{rescaled0}
\langle\tilde\Phi_{\D}'({\tilde \eta_1}, \vec x_1)\,\tilde\Phi_{\D}'(\tilde \eta_2, \vec x_2)\rangle= 
  \pi^{\frac{d}{2}} \frac{\Gamma\(\D- \frac{d}{2}\)}{\Gamma (\D)} \frac{\delta^{d}(\vec x_1 - \vec x_2)}{|\tilde\eta_{12}|^{2\D-d}} , \ \D> \frac{d}{2}
\ee
The Carrollian CFT$_{d+1}$ in which the two point function \eqref{rescaled0} is now defined lives on the ``flat'' manifold 
\be\label{flat1}
ds^2 = 0\cdot d\tilde\eta^2 + d\vec x^2
\ee

We now discuss the possible physical interpretation of the bound on the scaling dimension $\D$. 
 
\subsubsection{$\D>\frac{d}{2}$ as unitarity bound in the electric sector}

As $\D\rightarrow \frac{d}{2} +$ the finite part of the correlation function \eqref{electric} can be written as 
\be
\langle\tilde\Phi_{\D}({\tilde \eta_1}, \vec x_1)\,\tilde\Phi_{\D}(\tilde \eta_2, \vec x_2)\rangle = - \frac{\pi^{\frac{d}{2}}}{\Gamma(\frac{d}{2})} \ \tilde\eta_1^{\frac{d}{2}}\tilde\eta_2^{\frac{d}{2}} \  \ln |\tilde\eta_1 - \tilde\eta_2|^2 \ \delta^d(\vec x_1 - \vec x_2)
\ee
So in the electric sector of a Carrollian CFT$_{d+1}$, the scalar operator with dimension $\D = \frac{d}{2}$ is not a conformal primary. 
This is more obvious in terms of the rescaled correlation function which becomes 
\be
\langle\tilde\Phi_{\D}'({\tilde \eta_1}, \vec x_1)\,\tilde\Phi_{\D}'(\tilde \eta_2, \vec x_2)\rangle = - \frac{\pi^{\frac{d}{2}}}{\Gamma(\frac{d}{2})}  \  \ln |\tilde\eta_1 - \tilde\eta_2|^2 \ \delta^d(\vec x_1 - \vec x_2)
\ee
The situation is similar to that in two dimensions where a free scalar field with dimension zero is not a conformal field because it has logarithmic two point function. However, the derivative of the scalar field is a conformal primary operator. This was also observed in \cite{Mason:2023mti}.

In the present case the $\tilde\eta$ dependent part of the two point function \eqref{rescaled0} transforms like the correlation function of conformal primary operators of \textit{effective dimension $\D_{\t{eff}} = \D - \frac{d}{2}$}. $\D_{\t{eff}}$ vanishes at $\D= \frac{d}{2}$ and this leads to the logarithmic two point function. It is also easy to see that the derivatives $\partial_{\tilde\eta}^p \tilde\Phi_{d/2}'$ transform like Carrollian conformal fields of dimension $\D = \frac{d}{2} + p$ and has power-law correlation functions. All of these suggest that the $\D > \frac{d}{2}$ bound should be thought of as the unitarity bound. To be more precise, we suggest that, \\

\textit{In a \underline{unitary} Carrollian CFT$_{d+1}$ the dimension $\D$ of a scalar primary in the \underline{electric sector} is bounded from below by $\frac{d}{2}$, i.e, $\D > \frac{d}{2}$.} \\

In a relativistic CFT$_{d+1}$ the corresponding bound $\D \ge \frac{d-1}{2}$ is less stringent. 

It will be very interesting to prove the proposed unitarity bound from first principle. 

\subsubsection{Transformation of $\tilde\Phi_{\D}'$ under the Poincare group $\t{ISO}(3,1)$ }
In this section we show that $\tilde\Phi_{\D}'$ transforms like a Carrollian conformal primary under the near boundary (conformal) symmetry group ISO$(3,1)$ of EAdS$_3$. 

In order to find the transformation law of $\tilde\Phi_{\D}'(\tilde\eta,z,\bar z)$ we start with the transformation law 
\begin{align}
	\delta \Phi_{\D}(\eta,z, \bar z)=-\xi^{\mu}\,\partial_{\mu}\Phi_{\D}(\eta,z, \bar z)-\frac{\Delta}{3}\,\nabla_{\mu}\xi^{\mu}\,\Phi_{\D}(\eta,z,\bar z) \label{conf_primary}
\end{align}
of the (relativistic) scalar conformal primary $\Phi_{\D}(\eta, z, \bar z)$ under the Euclidean conformal group $\t{SO}(4,1)$. The vector field $\xi$ is a linear combination of the CKVs \eqref{so23_vec},
\begin{align}
	\xi=a_m \mathcal L_m+ \bar a_m \mathcal{\bar L}_m +p_{r,s}\,\mathcal P_{r,s}
\end{align} 

If the conformal symmetry is generated by the charge $Q_{\xi}$, then the commutator of the charge with the scalar primary $\Phi_{\D}(\eta,z, \bar z)$ should reproduce the transformation law $\eqref{conf_primary}$ of the field, i.e,
\begin{align}
	\left[Q_{\xi}, \Phi_{\D}(\eta,z, \bar z)\right]=-\xi^{\mu}\,\partial_{\mu}\Phi_{\D}(\eta,z, \bar z)-\frac{\Delta}{3}\,\nabla_{\mu}\xi^{\mu}\,\Phi_{\D}(\eta,z, \bar z).
\end{align}
This leads to the following commutation relation between the generators of $\t{SO}(4,1)$ and $\Phi_{\D}(\eta,z, \bar z)$,
\begin{align}
	&\left[\mathcal L_{m}, \Phi_{\D}\right]=z^{m+1}\,\partial_z \Phi_{\D}+\frac{1}{2}(m+1)\,z^m\,\eta\,\partial_{\eta}\Phi_{\D}-\frac{1}{2}\,m(m+1)\,\eta^2\,\partial_{\bar z}\Phi_{\D} \cr
	&\left[\mathcal{\bar L}_{m}, \Phi_{\D}\right]=\bar z^{m+1}\,\partial_{\bar z}\Phi_{\D}+\frac{1}{2}(m+1)\,\bar z^m\,\eta\,\partial_{\eta}\Phi_{\D}-\frac{1}{2}\,m(m+1)\,\eta^2\,\partial_{ z}\Phi_{\D}\cr
	&\left[\mathcal P_{r,s}, \Phi_{\D}\right]=z^{r+\frac{1}{2}}\,\bar z^{s+\frac{1}{2}}\,\left(\partial_{\eta}\Phi_{\D}-\frac{\Delta \Phi_{\D}}{\eta}\right)-\bigg(2\,\eta\,\left(s+\frac{1}{2}\right)\,z^{r+s}\partial_z\Phi_{\D}+2\,\eta\,\left(r+\frac{1}{2}\right)\,\bar z^{r+s}\partial_{\bar z}\Phi_{\D} \cr &+\left((r+\frac{1}{2})(s+\frac{1}{2})\right)\,\left(\eta^2\partial_{\eta}\Phi_{\D}+\Delta\,\eta\,\Phi_{\D}\right)\bigg). \label{so23_comm_Phi}
\end{align}
Now obtaining the (conformal Carrollian or) near-boundary limit of the transformation laws \eqref{so23_comm_Phi} is straightforward. We substitute \eqref{coord_scaling} and \eqref{flat} in \eqref{so23_comm_Phi} and take $\epsilon\rightarrow 0$. In this limit we get
\begin{align}
	&\left[ L_{m}, \tilde{\Phi}'_{\D}(\tilde \eta,z,\bar z)\right]=z^{m+1}\,\partial_z \tilde{\Phi}_{\D}'(\tilde \eta,z,\bar z)+(m+1)\,z^m\,\left(\frac{\tilde\eta}{2}\,\partial_{\tilde \eta}+\frac{\Delta}{2}\right)\tilde{\Phi}_{\D}'(\tilde \eta,z,\bar z) \cr
	&\left[\bar { L}_{m}, \tilde{\Phi}_{\D}'(\tilde \eta,z,\bar z)\right]=\bar z^{m+1}\,\partial_{\bar z}\tilde{\Phi}_{\D}'(\tilde \eta,z,\bar z)+(m+1)\,\bar z^m\,\left(\frac{\tilde \eta}{2}\,\partial_{\tilde \eta}+\frac{\Delta}{2}\right)\tilde{\Phi}_{\D}'(\tilde \eta,z,\bar z)\cr
	&\left[ P_{r,s}, \tilde{\Phi}_{\D}'(\tilde \eta,z,\bar z)\right]=z^{r+\frac{1}{2}}\,\bar z^{s+\frac{1}{2}}\,\partial_{\tilde \eta}\,\tilde{\Phi}_{\D}'(\tilde \eta,z,\bar z) \label{iso13_comm_phi}
\end{align}
where we have taken into account the scaling \eqref{trans} and also the relations \eqref{L}. 
Equation \eqref{iso13_comm_phi} is precisely the transformation law of a scalar Carrollian primary of dimension $\D$ \cite{Banerjee:2020kaa,Bagchi:2016bcd} in a Carrollian $\t{CFT}_3$ which lives on the manifold with degenerate metric \eqref{flat1}. Therefore near-boundary limit indeed contracts representations. 

\subsection{Magnetic Sector : \textbf{$\frac{d-1}{2} \le \D \le \frac{d}{2}$}} 
When the conformal dimension $\D$ of the (relativistic) conformal field is $< \frac{d}{2}$, the second term in the expansion \eqref{near} dominates as we take $\epsilon\rightarrow 0$. So in the near boundary limit we get 
\be
\langle\Phi_{\D}(\epsilon\,{\tilde \eta_1}, \vec x_1)\,\Phi_{\D}(\epsilon\, \tilde \eta_2, \vec x_2)\rangle_{\t{Near Boundary}} \\
\rightarrow  \epsilon^{2\D} \(\frac{\tilde\eta_1^{\D}\tilde\eta_2^{\D}}{|\vec x_1 - \vec x_2|^{2\D}}\), \  \frac{d-1}{2} \le \D \le \frac{d}{2}
\ee
The lower bound on $\D$ is nothing but the unitarity bound of the parent relativistic CFT$_{d+1}$. Now we can define the rescaled field 
\be
\Phi_{\D}(\epsilon\tilde\eta, \vec x) = \epsilon^{\D} \tilde\Phi_{\D}(\tilde\eta, \vec x)
\ee
in terms of which the two point function becomes 
\be\label{mag}
\langle\tilde\Phi_{\D}({\tilde \eta_1}, \vec x_1)\,\tilde\Phi_{\D}(\tilde \eta_2, \vec x_2)\rangle_{\t{Near Boundary}} = \frac{\tilde\eta_1^{\D}\tilde\eta_2^{\D}}{|\vec x_1 - \vec x_2|^{2\D}}, \  \frac{d-1}{2} \le \D \le \frac{d}{2}
\ee
This is the two point function of Carrollian conformal primaries in the magnetic sector. Equation \eqref{mag} suggests the following unitarity bound \\

\textit{In a \underline{unitary} Carrollian CFT$_{d+1}$ the dimension $\D$ of a scalar primary in the \underline{magnetic sector} is both bounded from below and above by $\frac{d-1}{2} \le \D \le \frac{d}{2}$.} \\

The lower limit is simply the unitarity bound of the parent relativistic CFT. The upper bound is more mysterious. It will be very interesting to see how this can be derived from the first principle. \\

To summarize, in the near boundary limit a scalar conformal primary of dimension $\D$ in a relativistic CFT$_{d+1}$ becomes electric Carrollian primary if $\D> \frac{d}{2}$ and becomes magnetic Carrollian primary if $\frac{d-1}{2} \le \D \le \frac{d}{2}$.

\section{Near boundary limit of (non-conformal) field theory EAdS$_3$}

Let us consider a scalar field of mass $m$ on Euclidean AdS$_3$. We take the AdS$_3$ metric to be 
\be\label{Nads}
ds^2 = d\phi^2 + e^{2\phi} d\gamma d\bar\gamma
\ee
where $-\infty<\phi<\infty$ and $\gamma$ and $\bar\gamma$ are complex conjugates. In these coordinates the AdS boundary is at $\phi = \infty$. 

The relation between the coordinates $(\phi, \gamma, \bar\gamma)$ and the ones used in \eqref{ds3_planar} is
\be
\phi = -\ln\eta, \  \gamma = z, \  \bar\gamma = \bar z
\ee
In terms of $\phi$ the near boundary limit \eqref{coord_scaling} becomes
\be\label{coor}
\phi = - \ln\epsilon + \tilde\phi, \  \epsilon\rightarrow 0
\ee
with $\tilde\phi$ held fixed. 

The action of the scalar field is given by 
\be\label{scalar}
S = \frac{1}{2} \int d\phi d\gamma d\bar\gamma \ e^{2\phi} \( (\partial_{\phi} \Psi)^2 + e^{-2\phi} \partial_i\Psi \partial_i\Psi + m^2 \Psi^2  \)
\ee
We can see that in the near boundary limit $\phi\rightarrow\infty$, we can neglect the spatial derivative term and the action becomes 
\be\label{rescaled}
S_{\t{NB}} = \frac{1}{2} \int d\phi d\gamma d\bar\gamma \ e^{2\phi} \( (\partial_{\phi} \Psi)^2 + m^2 \Psi^2  \)
\ee
The procedure can be formalized as follows. First we make the change of coordinate \eqref{coor} and do the field rescaling \eqref{elec} $\Psi\rightarrow \epsilon\tilde\Psi$. This gives us 
\be
S = \frac{1}{2} \int d\tilde\phi d\gamma d\bar\gamma \ e^{2\tilde\phi} \( (\partial_{\tilde\phi} \tilde\Psi)^2 + \epsilon^2 e^{-2\tilde\phi} \partial_i\tilde\Psi \partial_i\tilde\Psi + m^2 \tilde\Psi^2  \)
\ee
Now we take $\epsilon\rightarrow 0$ keeping all other quantities fixed and we get 
\be\label{nbf}
S_{\t{NB}} = \frac{1}{2} \int d\tilde\phi d\gamma d\bar\gamma \ e^{2\tilde\phi} \( (\partial_{\tilde\phi} \tilde\Psi)^2 + m^2 \tilde\Psi^2  \)
\ee 
Note that here the range of $\tilde\phi$ is again $(-\infty, \infty)$. So as expected \eqref{nbf} and \eqref{rescaled} have identical structure except the replacement $(\phi, \Psi)\rightarrow (\tilde\phi, \tilde\Psi)$. \\

So for notational simplicity we will work with the action \eqref{rescaled} \textit{with the understanding that the range of $\phi$ is the whole real line $(-\infty, \infty)$}.

\subsection{(Infinite) Symmetries of the near-boundary action}

The near boundary action 
\be\label{nba}
S_{\t{NB}} = \frac{1}{2} \int d\phi d\gamma d\bar\gamma \ e^{2\phi} \( (\partial_{\phi} \Psi)^2 + m^2 \Psi^2  \)
\ee
is invariant under SO$(3,1)$ which acts on the field and coordinates as follows 
\be
\begin{gathered}
\Psi'(\phi', \gamma', \bar\gamma') = \Psi(\phi, \gamma, \bar\gamma) \\
\phi' = \phi + \ln|c\gamma + d|^2, \  
\gamma' = \frac{a\gamma + b}{c\gamma + d}, \   \bar\gamma' = \frac{\bar a\bar\gamma + \bar b}{\bar c\bar \gamma + \bar d}, \  
\begin{pmatrix}
a &b \\
c& d 
\end{pmatrix} \in SL(2,\mathbb{C})
\end{gathered}
\ee
The coordinate transformation is the isometry of the degenerate metric 
\be\label{deg}
ds^2 = e^{2\phi} d\gamma d\bar\gamma 
\ee 
on which the scalar field lives. However, the metric \eqref{deg} and the action \eqref{nba} has the infinite dimensional local symmetry of arbitrary holomorphic reparametrization given by 
\be
\begin{gathered}
\Psi'(\phi', \gamma', \bar\gamma') = \Psi(\phi, \gamma, \bar\gamma) \\
\phi' = \phi - \frac{1}{2} \ln\bigg|\frac{d\gamma'}{d\gamma}\bigg|^2, \  
\gamma' = \gamma'(\gamma), \   \bar\gamma' = \bar\gamma'(\bar\gamma) \  
\end{gathered}
\ee
Therefore in the near boundary limit in AdS$_3$ the symmetry of the bulk action is enhanced to the infinite dimensional Virasoro algebra. 

\subsection{Boundary correlation function}
We want to compute the boundary two point correlation function when the bulk action is given by 
\be
S_{\t{NB}} = \frac{1}{2} \int d\phi d\gamma d\bar\gamma \ e^{2\phi} \( (\partial_{\phi} \Psi)^2 + m^2 \Psi^2  \)
\ee
The boundary is at $\phi=\infty$. To compute the boundary correlation function we first compute the bulk Green's function defined as 
\be
\( - e^{-2\phi}\frac{\partial}{\partial\phi}\( e^{2\phi} \frac{\partial}{\partial\phi}\) + m^2 \) G(\phi, \gamma, \bar\gamma; \phi', \gamma', \bar\gamma') = \frac{\delta(\phi - \phi') \delta^2(\gamma - \gamma')}{e^{2\phi}}
\ee
The Green's function $G$ can be easily computed to be 
\be
\begin{gathered}
G(\phi, \gamma, \bar\gamma; \phi', \gamma', \bar\gamma') \\ = 
\frac{e^{- \D_{+} \phi} e^{-{\D_{-}\phi'}}}{2\sqrt{1+m^2}} \theta(\phi - \phi') \delta^2(\gamma - \gamma') + \frac{e^{- \D_{-} \phi} e^{-{\D_{+}\phi'}}}{2\sqrt{1+m^2}} \theta(\phi' - \phi) \delta^2(\gamma - \gamma')
\end{gathered}
\ee
where 
\be
\D_{\pm} = 1 \pm \sqrt{1+m^2}
\ee
The form of the Green's function becomes more familiar if we write it in terms of the coordinate $\eta = e^{-\phi}$. The boundary is now located at $\eta=0$. In term of $\eta$ we get
\be
\begin{gathered}
G(\eta, \gamma, \bar\gamma; \eta', \gamma', \bar\gamma') \\ = 
\frac{{\eta}^{\D_{+}} \eta'^{\D_{-}}}{2\sqrt{1+m^2}} \theta(\eta' - \eta) \delta^2(\gamma - \gamma') + \frac{{\eta}^{\D_{-}} \eta'^{\D_{+}}}{2\sqrt{1+m^2}} \theta(\eta - \eta') \delta^2(\gamma - \gamma')\end{gathered}
\ee
Now we can define the boundary two point function using the \textit{extrapolate} dictionary.  So we take $(\eta,\eta')$ to zero and the boundary correlator we get is 
\be\label{bc}
\begin{gathered}
\langle{O_{\D_+}(\gamma, \bar\gamma) O_{\D_{-}}(\gamma',\bar\gamma')}\rangle = \delta^2(\gamma - \gamma'), \    \D_+ + \D_- = 2 \\
\langle{O_{\D_+}(\gamma, \bar\gamma) O_{\D_{+}}(\gamma',\bar\gamma')}\rangle = \langle{O_{\D_-}(\gamma, \bar\gamma) O_{\D_{-}}(\gamma',\bar\gamma')}\rangle = 0
\end{gathered}
\ee
So the boundary two point function is a pure contact term because the the theory is ultralocal in the boundary direction. 
It is easy to check that \eqref{bc} is conformally invariant. The dimensions of the boundary operators is fixed by the exponents of the $(\eta,\eta')$. So we conclude that \\

\textit{In the near boundary limit in AdS$_{d+1}$ a scalar field of mass $m$ corresponds to two boundary operators of dimensions} 
\be\nonumber
\D_{\pm} = \frac{d}{2} \pm \sqrt{\frac{d^2}{4} + m^2}
\ee

\section{Near boundary limit of the Worldline action}\label{wl}
In this section we study the worldline action of a particle of mass $m^2$ moving on the Euclidean AdS$_3$. Our goal is to take the near boundary limit in the worldline action. 

With usual notation the worldline action can be written as
\be
S = \frac{1}{2} \int d\lambda \( \frac{1}{e(\lambda)} g_{ab} \frac{dx^a}{d\lambda} \frac{dx^b}{d\lambda} - m^2 e(\lambda)\)
\ee
where $g_{ab}$ is the AdS$_3$ metric \eqref{Nads}.

We now choose the standard gauge 
\be
e(\lambda) = 1
\ee
In this gauge the action becomes 
\be
\begin{gathered}
S = \frac{1}{2} \int d\lambda \(g_{ab} \frac{dx^a}{d\lambda} \frac{dx^b}{d\lambda} - m^2 \) 
= \frac{1}{2} \int d\lambda \(\(\frac{d\phi}{d\lambda}\)^2 + 2 e^{2\phi} \frac{d\gamma}{d\lambda}\frac{d\bar\gamma}{d\lambda} - m^2 \)
\end{gathered}
\ee
To take the near boundary limit we have to take $\phi\rightarrow\infty$. In order to take this limit we introduce two Lagrange multiplier fields $(\beta(\lambda), \bar\beta(\lambda))$ and write the action as,
\be\label{first}
\begin{gathered}
S 
= \int d\lambda \(\frac{1}{2}\(\frac{d\phi}{d\lambda}\)^2 + \beta \frac{d\gamma}{d\lambda} + \bar\beta \frac{d\bar\gamma}{d\lambda} - \frac{m^2}{2} - \beta \bar\beta e^{-2\phi} \)
\end{gathered}
\ee
In the $\phi\rightarrow\infty$ limit we can neglect the last term in \eqref{first} and get
\be\label{first2}
\begin{gathered}
S_{\t{NB}} 
= \int d\lambda \(\frac{1}{2}\(\frac{d\phi}{d\lambda}\)^2 + \beta \frac{d\gamma}{d\lambda} + \bar\beta \frac{d\bar\gamma}{d\lambda} - \frac{m^2}{2}\)
\end{gathered}
\ee
Again we take $-\infty<\phi<\infty$ in \eqref{first2}.

The Hamiltonian derived from the near-boundary action \eqref{first2} is given by 
\be
H = \frac{1}{2} \( p_{\phi}^2 + m^2\)
\ee
where $p_{\phi}$ is the canonical momentum conjugate to $\phi$. Now if we quantize the worldline theory then the physical states satisfy the constraint (or the missing equation of motion for $e(\lambda)$) 
\be
H\ket{\t{phys}} = 0 
\ee
So the wave functions $\tilde\Psi(\phi, \gamma, \bar\gamma)$ satisfy the equation 
\be\label{1q}
\frac{1}{2}\( - \frac{\partial^2}{\partial\phi^2} + m^2\) \tilde\Psi(\phi,\gamma,\bar\gamma) = 0
\ee
where we have substituted, $p_{\phi} = -i \frac{\partial}{\partial\phi}$. The Schrodinger equation \eqref{1q} can be derived from the space-time action 
\be
\tilde S = \frac{1}{2} \int d\phi d\gamma d\bar\gamma \( (\p_{\phi}\tilde\Psi)^2 + m^2 {\tilde\Psi}^2\)
\ee
To bring it to the more familiar form we define 
\be
\Psi = e^{-\phi} \tilde\Psi
\ee
in terms of which the action becomes 
\be\label{1q2}
S = \frac{1}{2} \int d\phi d\gamma d\bar\gamma \ e^{2\phi} \( (\p_{\phi}\Psi)^2 + m^2 {\Psi}^2\)
\ee
The action \eqref{1q2} matches precisely with the action \eqref{rescaled} which is the near boundary limit of the action of a massive scalar field theory on Euclidean AdS$_3$. \\

\textit{So if we take the near boundary limit of the worldline action, written in the first order form, then the space-time theory we get (after first quantization) is an \underline{electric Carroll} theory of a massive scalar field.}  \\

We will use this result to understand the near-boundary limit of the string sigma model of Euclidean AdS$_3$. Before we end this section let us discuss the physical picture. 

The equations of motion of the fields $(\beta, \bar\beta)$ following from the action \eqref{first2} are 
\be
\gamma = \t{const}, \  \bar\gamma = \t{const}
\ee
So the particle described by \eqref{first2} is pinned at a point of the boundary and moves only in the radial direction in the bulk. This is why the near boundary limit of the action of a massive scalar field \eqref{rescaled} does not have any spatial derivatives and is described by an ultralocal (electric) Carrollian theory. 

This worldline picture will be useful when we study the near boundary limit of string theory on EAdS$_3$ in the presence of NS-NS two form flux.

\section{Near boundary limit of string theory on AdS$_3$ and the dual celestial CFT$_2$}\label{7}
The conventional string theories do not have target space conformal invariance and in the low energy limit give Einstein gravity rather than conformal gravity. Therefore in the near boundary limit string theories on AdS should be dual to Celestial CFTs with only conformal invariance but not Poincare invariance \footnote{To study Celestial CFT$_2$ with ISO$(3,1)$ invariance one needs to study the near boundary limit of \textit{Twistor String Theory} on AdS$_3$. This is beyond the scope of this paper.}. 

In this section we consider bosonic string theory on Euclidean AdS$_3\times \mathcal N$ in the presence of Neveu-Schwarz $B_{\mu\nu}$ field and $\mathcal{N}$ is some compact space. The AdS$_3$ metric is
\be
ds^2 = l^2 \( d\phi^2 + e^{2\phi} d\gamma d\bar\gamma\)
\ee 
where $l$ is the AdS$_3$ radius.

Now the bosonic part of the worldsheet Lagrangian on AdS$_3$ is given by the Wess-Zumino-Witten model for the coset $SL(2,\mathbb{C})/ SU(2)$ and can be written as \cite{Giveon:1998ns,deBoer:1998gyt,Maldacena:2000hw}
\be
S = \frac{2l^2}{l_{s}^2}\int d^2w \( \partial\phi \bar\partial\phi + e^{2\phi} \bar\partial\gamma \partial\bar\gamma\) + S_{\t{int}}
\ee
where $S_{\t{int}}$ is the action describing string propagation on compact internal space. 

Now to take the near boundary limit we write the action in the first order form as \cite{Giveon:1998ns,deBoer:1998gyt,Maldacena:2000hw} 
\be\label{fo}
S = \int d^2w \( \partial\phi \bar\partial\phi - \frac{2}{\alpha_{+}} \sqrt{g} R \phi + \beta\bar\partial\gamma + \bar\beta\partial\bar\gamma - \beta\bar\beta e^{-\frac{2}{\alpha_{+}}\phi} \) + S_{\t{int}}
\ee
where $(\beta,\gamma)$ are bosonic holomorphic fields with weights $(1,0)$ and $(0,0)$. $(\bar\beta,\bar\gamma)$ are the corresponding antiholomorphic fields. The parameter $\alpha_{+}$ is related to the string length $l_s$ and the AdS radius $l$ by
\be
\alpha_{+}^2 = 2k -4 = 2 \frac{l^2}{l_s^2} - 4
\ee 
$R$ is the worldsheet curvature. 

In the near boundary limit $\phi\rightarrow\infty$ the last term in the action \eqref{fo} can be neglected and we get the free worldsheet action, 
\be\label{action1}
\begin{gathered}
S_{NB} = \int d^2w \( \partial\phi \bar\partial\phi - \frac{2}{\alpha_{+}} \sqrt{g} R \phi + \beta\bar\partial\gamma + \bar\beta\partial\bar\gamma \) + S_{\t{int}}
\end{gathered}
\ee
That this is the correct near boundary limit follows from our analysis of the near boundary limit of the worldline action in section-\eqref{wl}.

\subsection{Properties of the near boundary worldsheet action}
The free worldsheet action
\be\label{action2}
 S_{NB} = \int d^2w \( \partial\phi \bar\partial\phi - \frac{2}{\alpha_{+}} \sqrt{g} R \phi + \beta\bar\partial\gamma + \bar\beta\partial\bar\gamma \) + S_{\t{int}}
\ee
has been studied in the context of the AdS$_3$-CFT$_2$ correspondence and below we summarize some of the key points:  

1) The equation of motion of $\beta$($\bar\beta$)
\be\nonumber
\bar\partial\gamma = 0, \  \partial\bar\gamma = 0
\ee
require $\gamma$ ($\bar\gamma$) to be a holomorphic (anti-holomorphic) map from the worldsheet to the boundary sphere\footnote{We are calling it boundary sphere because we have already taken the near boundary limit.} ($S^2$) in the target space. These worldsheet instantons or wrapped strings are known as the ``\textit{long-strings}'' \cite{Giveon:1998ns,deBoer:1998gyt,Seiberg:1999xz,Maldacena:2000hw}. 

2) The dual or space-time CFT$_2$ describing the long-strings has central charge \cite{Giveon:1998ns} $c=6kp$ where $p$ is the number of times the worldsheet wraps the boundary sphere. Seiberg and Witten \cite{Seiberg:1999xz} argued that the space-time CFT$_2$ which describes the long strings has a Liouville sector with background charge (for $p=1$)
\be
Q = (k-3) \sqrt{\frac{2}{k-2}}
\ee
and central charge\footnote{$c_{\t{L}} = 6k - c_{\t{int}}$ where $c_{\t{int}}$ is the central charge of $S_{\t{int}}$ which describes the motion of the string in the $23$ dimensional compact space. $c_{\t{int}}$ can be determined in terms of $k$ using the condition that the total matter central charge of the worldsheet CFT is $26$. } 
\be
c_{\t{L}}= 1+ 3 Q^2 = 1+ 6 \frac{(k-3)^2}{k-2}
\ee
The Liouville field describes the radial ($\phi$) fluctuations of the long-string worldsheet. 

3) It was shown in \cite{Giveon:1998ns} that the affine symmetries of the free action \eqref{action2} can be ``lifted" to the space-time and become affine symmetries of the dual CFT$_2$. To be more precise, if the worldsheet theory has an affine $\hat G$ symmetry with level $k'$ then after lifting to the space time it becomes an affine $\hat G$ symmetry of the dual CFT$_2$ with level $\tilde k$ given by $\tilde k = p k$. Here $p$ is the number of times the string worldsheet wraps the boundary sphere. 

Similarly the affine $SL_2\times SL_2$ symmetry of the action \eqref{action2}, after lifting to the space time, becomes the Virasoro algebra of the dual CFT$_2$ with central charge $c= 6kp$. 

\subsection{Dual Celestial CFT$_2$}
We have seen that in the near boundary limit the worldsheet action reduces to 
\be
S_{NB} = \int d^2w \( \partial\phi \bar\partial\phi - \frac{2}{\alpha_{+}} \sqrt{g} R \phi + \beta\bar\partial\gamma + \bar\beta\partial\bar\gamma \) + S_{\t{int}}
\ee
According to our proposal: 

\textit{The space-time CFT$_2$ dual to the string theory described by the worldsheet action $S_{NB}$ is an example of a Celestial CFT$_2$ (with only Lorentz invariance).} The Celestial CFT$_2$ contains a \textit{Liouville} sector which describes the radial fluctuation of the long string worldsheet \footnote{A more precise statement is contained in \cite{Eberhardt:2019qcl}. See also \cite{Knighton:2024pqh} for recent work on long string CFT.}. Other fields which live on the long string worldsheet are the conformal fields which describe the position of the string in the compact space. Moreover, the infinite dimensional symmetries of the Celestial CFT$_2$ can be thought of as lifts of the infinite dimensional symmetries on the worldsheet. 

In very interesting recent works \footnote{See also \cite{Ogawa:2024nhx,Mol:2024qct} for related work.} \cite{Stieberger:2023fju,Melton:2024akx,Donnay:2025yoy} examples of Celestial CFT$_2$ were constructed which are Liouville theory coupled to various matter fields. These theories are Lorentz invariant but not Poincare invariant. The existence of the Liouville sector in these examples and in the long-string CFT$_2$ suggests that Liouville sector could be a universal feature of all Celestial CFT$_2$. 

We would also like to point out that the relation between the worldsheet CFT and the Celestial CFT, that we find in this paper, seems to resonate with the observations of \cite{Stieberger:2018edy,Kervyn:2025wsb} that in the high energy limit in asymptotically flat space-time one can identify the string worldsheet with the celestial sphere. It will be interesting to clarify the connection between our work and the works of \cite{Stieberger:2018edy,Kervyn:2025wsb}. 

Last but not the least, the techniques developed in \cite{Eberhardt:2019ywk} and the related papers, may help us to understand the nature of Celestial CFT$_2$ better. We hope to return to this in near future.

\section{Acknowledgement} 
We are grateful to Ashoke Sen for very helpful correspondence. We also like to thank Sujay Ashok, Nilay Kundu, Alok Laddha, Partha Paul, Nemani V. Suryanarayana and Sandip Trivedi for useful discussions. SB would like to thank all the participants of the Chennai String School for interesting questions related to this work. The work of SB and NG is supported by the  Swarnajayanti Fellowship (File No- SB/SJF/2021-22/14) of the Department of Science and Technology and SERB, India.

\begin{appendices} 

\section{Near boundary limit of Einstein Gravity on EAdS$_3$}
In this appendix, we derive the near boundary limit of Einstein-Hilbert action and the equations of motion.
For this let us consider the metric of locally EAdS$_3$ in Fefferman-Graham gauge with the following line element
\begin{align}
	ds^2=d\phi^2+e^{2\phi}h_{ij}(\phi, x^i)dx^i dx^j. \label{line_element}
\end{align}
The boundary coordinates are labelled by $x^i=\{\gamma, \bar \gamma\}$. On a constant $\phi$ hypersurface, the two dimensional induced boundary metric $a_{ij}$ is
\begin{align}
	a_{ij}= e^{2\phi} h_{ij}
\end{align} 
The Einstein-Hilbert action for EAdS$_3$ gravity is given by,
\begin{align}
	S_{EH}=\int d\phi d^2\gamma \sqrt{g}(R+2),
\end{align}
where we have set AdS radius $l=1$.
For the metric in \eqref{line_element}, the action becomes
\begin{align}
	S_{EH}=	\int d\phi\,d^2\gamma\,e^{2\phi}\, \sqrt{h}\left(K^2-K_{ij}K^{ij}+2\right)+ \int d\phi\, d^2\gamma  \sqrt{h} \mathcal\, \mathcal R^{(2)}, \label{EH_action}
\end{align}
where the extrinsic curvature $K_{ij}$ is defined as
\begin{align} 
	K_{ij}=\frac{1}{2}\partial_{\phi} a_{ij}, ~~K=a^{ij}\,K_{ij}.
\end{align} 
and $\mathcal R^{(2)}$ is the Ricci scalar of metric $h_{ij}$. One can see that in the near boundary limit $\phi \rightarrow \infty $ , the first term in  \eqref{EH_action} dominates and therefore the near boundary action is given as \footnote{ Note that the procedure to obtain near boundary limit that was formalised in the case of scalar field theory will lead to the same expression in terms of $\tilde{\phi}$ but for notational simplicity we work with the coordinate $\phi$.}
\begin{align}
	S_{NB}=\int d\phi\,d^2\gamma\,e^{2 \phi}\, \sqrt{h}\left(K^2-K_{ij}K^{ij}+2\right) \label{carr_action}, \  -\infty< \phi < \infty	.
\end{align} 
This matches with the Carrollian limit of the Einstein-Hilbert action derived in \cite{Hansen:2021fxi}.

The action \eqref{carr_action} is
defined on the degenerate manifold with line element
\begin{align}
	ds^2= 0. d \phi^2+ e^{2\phi} h_{ij}(\phi,x^i)dx^i dx^j.
\end{align}  
The Einstein equation is,
\begin{align}
	G_{\mu \nu}=	R_{\mu \nu}-\frac{1}{2}g_{\mu \nu}-g_{\mu \nu}=0 \label{eom}
\end{align}
Since we already fixed the gauge to be \eqref{line_element}, $G_{\phi \phi}$ and $G_{\phi i}$ component of the equations of motion \eqref{eom} are constraints given as,
\begin{align}
	&G_{\phi \phi}=\frac{1}{2}\left(K^2-K_{ij}K^{ij}-2\right)-\frac{1}{2}e^{-2 \phi}\,\mathcal R^{(2)}=0 \label{cons1} \end{align}
\begin{align}
	&G_{\phi i}=\left(a^{kl}\nabla_{k}K_{il}-a^{kl}\nabla_{i}K_{kl}\right)=0 \label{cons2}
\end{align}
The covariant derivative $\nabla$ is with respect to $a_{ij}$. The $G_{ij}$ component is 
\begin{align}
	G_{ij}=\left(a_{ij}\,\partial_{ \phi}K+K_{mn}K^{mn}\,a_{ij}-a_{ij}\right)-\partial_{ \phi}K_{ij}+a^{nl }K_{i l}\,K_{nj}=0 \label{cons3}
\end{align} 
In the limit $\phi \rightarrow \infty$, the equation \eqref{cons1} becomes,
\begin{align}
	K^2-K_{ij}K^{ij}-2=0 \label{cons4}
\end{align}
whereas the equations for component $G_{\phi i}$ and $G_{ij}$ remain unchanged as all the terms appear at the same order in $e^{2\phi}$.

To write the equations in term of $h_{ij}$, we introduce the following quantities
\begin{align}
	\zeta_{ij}=\frac{1}{2}\partial_{ \phi}h_{ij}\,,~~\Theta=h^{ij}\zeta_{ij}
\end{align}
then the equation \eqref{cons4}, \eqref{cons2}, \eqref{cons3} are given by,
\begin{align}
	\Theta-\frac{1}{2}\zeta_{ij}\zeta^{ij}+\frac{1}{2}\Theta^2=0 \label{cons11}
\end{align}
\begin{align}
	h^{kl}\nabla_{k}\zeta_{il}-h^{kl}\nabla_{i}\zeta_{kl}=0 \label{cons21}
\end{align}
\begin{align}
	\left(\partial_{\tilde \phi}\Theta\,h_{ij}-\partial_{\tilde \phi}\zeta_{ij}+\zeta_{kl}\zeta^{kl}\,h_{ij}+\zeta_{i}^{k}\zeta_{kj}+2\Theta h_{ij}-2 \zeta_{ij}\right)=0 \label{EOM1}
\end{align}
All indices are raised and lowered by $h_{ij}$.
\subsection{Solutions to the equations}
Taking trace of \eqref{EOM1} we have,
\begin{align}
	2 \Theta+\partial_{ \phi}\Theta+ \zeta_{ij}\zeta^{ij}=0  \label{treq}.
\end{align}
One can use \eqref{treq} to  simplify \eqref{EOM1} as follows,
\begin{align}
	\partial_{ \phi}\zeta_{ij}+2\zeta_{ij}-\zeta_{i}^{k}\zeta_{kj}=0
\end{align}
The solution of above equation is exact and can be written as,
\begin{align}
	h_{ij}( \phi, x^i)= h^{(0)}_{ij}(x^i)+e^{-2  \phi} h^{(1)}_{ij}(x^i)+e^{-4  \phi} h^{(2)}_{ij}(x^i) \label{series_exp}
\end{align}
with $h^{(2)}_{ij}(x^i)$ given by,
\begin{align}
	h^{(2)}_{ij}=\frac{1}{4} h^{(1)}_{ik}h_{(0)}^{kl}h^{(1)}_{lj}
\end{align}
Substituting \eqref{series_exp} in \eqref{cons11}, $\mathcal O(e^{-2 \phi})$ component implies,
\begin{align}
	\boxed{h^{ij}_{(0)}\,h^{(1)}_{ij}=0.} \label{traceless}
\end{align}
The equation \eqref{cons21} to the lowest order gives
\begin{align}
	-h_{(0)}^{kl}D_{k}h^{(1)}_{il}+ h_{(0)}^{kl}D_ih^{(1)}_{kl}=0
\end{align}
where $D_i$ is covariant derivative  with respect to $h^{(0)}_{ij}$. Due to the trace condition second term vanishes and therefore, one gets the conservation law,
\begin{align}
	\boxed{D^k\,h^{(1)}_{ki}=0.} \label{cons_eq}
\end{align}
To summarize, the solution space is parameterized by $h^{(0)}_{ij}$ and $h^{(1)}_{ij}$ with the latter being a conserved and traceless tensor.
\paragraph{Comments:} The tracelessness of $h^{(1)}_{ij}$ in \eqref{traceless} implies that the \textit{Celestial CFT$_2$ dual to the near boundary limit of Einstein gravity has zero central charge.}  

From \eqref{cons_eq},  the boundary stress tensor of this dual Celestial CFT$_2$ will have holomorphic and anti-holomorphic components that will generate two commuting copies of Virasoro algebra.  One can contrast this with conventional AdS$_3$/CFT$_2$ setup where the symmetry algebra of boundary CFT is two commuting copies of Virasoro algebra with \textit{non-zero} central charge \cite{Brown:1986nw}.

\subsection{Extension to higher dimension EAdS$_{d+1}$}
Here we extend the results of  previous section to general $d+1$ dimensions. The Einstein-Hilbert action in the Fefferman-Graham gauge becomes, 
\begin{align}
	S_{EH}=	\int d \phi\,d^{d}x\,e^{d \phi}\, \sqrt{h}\left(K^2-K_{ij}K^{ij}-2\Lambda\right)+ \int d\phi \,d^d x  e^{(d-2)\phi}\sqrt{h} \mathcal\, \mathcal R^{(d)}
\end{align}
As $\phi \rightarrow \infty$, the action is given by,
\begin{align}
	S_{NB}=	\int d \phi\,d^{d}x\,e^{d  \phi}\, \sqrt{h}\left(K^2-K_{ij}K^{ij}-2\Lambda\right)
\end{align}
The constraint equations in the near boundary limit is given by,
\begin{align}
	\frac{1}{2}\left(K^2-K_{ij}K^{ij}+2 \Lambda\right)=0
\end{align}
\begin{align}
	\left(a^{kl}\nabla_{k}K_{il}-a^{kl}\nabla_{i}K_{kl}\right)=0 
\end{align}
The equation of motion is given by,
\begin{align}
	-K\,K_{ij}+\frac{1}{2}K_{mn}K^{mn}\,a_{ij}+\Lambda\,a_{ij}- a_{in}\,\partial_{ \phi}K^{n}_{~j}+\partial_{ \phi} K\,a_{ij}+\frac{1}{2} K^2\,a_{ij}=0
\end{align}
The constraint equations in terms of the  metric $h_{ij}$ are,
\begin{align}
	&\Theta^2-\zeta_{ij}\zeta^{ij}+(d-1)(d+2\Theta)+2\Lambda=0 \cr
	& 	h^{kl}\nabla_{k}\zeta_{il}-h^{kl}\nabla_{i}\zeta_{kl}=0	
\end{align}
The equation for $h_{ij}$ can be written as,
\begin{align}
	\frac{h_{ij}}{2}\left(\zeta_{mn}\zeta^{mn}+2\partial_{\phi}\Theta+\Theta^2+2\Theta\,d+d(d-1)\right)-(\Theta+d)\zeta_{ij}-a_{in}\partial_{ \phi}\zeta^n_{~j}=0
\end{align}

\end{appendices}

\bibliographystyle{utphys}
\providecommand{\href}[2]{#2}\begingroup\raggedright

\end{document}